%% file: sec_0.tex
\documentclass[
reprint,
superscriptaddress,
nofootinbib,
aps,
pra
]{revtex4-2}

\usepackage{amsmath, amssymb, amsfonts}
\usepackage{physics}
\usepackage{graphicx}
\usepackage{subcaption}
\usepackage[percent]{overpic}
\usepackage{bm}
\usepackage{hyperref}
\usepackage{xcolor}
\usepackage[T2A]{fontenc}
\usepackage[utf8]{inputenc}
\usepackage{tikz}
\usepackage{quantikz}
\usepackage{ragged2e}
\newcommand{\trc}{\mathrm{Tr}}
\newcommand{\be}{\begin{equation}}
\newcommand{\ee}{\end{equation}}

\newcommand{\Hd}{d}                         % Hilbert-space dimension

\newcommand{\cG}{\mathcal{G}}      % implemented noisy PTM channel
\newcommand{\Tgt}{\mathcal{T}}     % ideal target PTM channel
\newcommand{\cU}{\mathcal{U}}      % unitary PTM channel

\newcommand{\Err}{E}               % total additive PTM error matrix

\newcommand{\Ec}{E_{c}}            % commuting additive PTM error matrix
\newcommand{\Ea}{E_{a}}            % anticommuting additive PTM error matrix

\newcommand{\Eu}{E_{u}}            % coherent/unitary additive PTM error matrix
\newcommand{\chE}{\mathcal{E}}   % total error compositional CPTP channel
\newcommand{\chEu}{\mathcal{E}_{u}}   % coherent error compositional CPTP channel
\newcommand{\chEv}{\mathcal{E}_{v}}   % dissipative error compositional CPTP channel

\newcommand{\chZ}{\mathcal{Z}}              % virtual-Z operation
\newcommand{\Fid}{F}                       % fidelity
\newcommand{\eF}{\varepsilon}              % infidelity

\newcommand{\acom}{a}                      % anticommutative scalar part
\newcommand{\com}{c}                       % commutative scalar part
\newcommand{\uni}{u}                       % unitary scalar part
\newcommand{\nuni}{v}                      % non-unitary scalar part

\newcommand{\pobs}{p}                      % measured probability
\newcommand{\wfeat}{w}                     % feature weight
\newcommand{\wzero}{w_0}                   % offset

\newcommand{\midx}{m}                      % feature index
\newcommand{\Mid}{M}                       % number of selected features
\newcommand{\ki}{k}                        % ensemble index
\newcommand{\Ke}{K}                        % ensemble size

\newcommand{\Ns}{N}                        % shots
\newcommand{\Lrep}{L}                      % repetition depth

\newcommand{\Iccz}{I_{\mathrm{CCZ}}}      % CCZ-derived identity surrogate

\newcommand{\RZfourth}[1]{%
\gate[style={fill=orange!15}]{R_Z(#1)}
}

\newcommand{\tp}{^{\mathsf T}}
\begin{document}

\title{Non-Clifford Benchmarking via Ensemble Feature Selection}

\author{Stancho G. Stanchev}
\affiliation{Center for Quantum Technologies, St Kliment Ohridski University of Sofia, 5 James Bourchier blvd, 1164 Sofia, Bulgaria}

\author{Nikolay V. Vitanov}
\affiliation{Center for Quantum Technologies, Faculty of Physics, St Kliment Ohridski University of Sofia, 5 James Bourchier blvd, 1164 Sofia, Bulgaria}

\date{\today}

\begin{abstract}
%A.1
We propose an Ensemble Feature Selection (EFS) method for
fast estimation of process infidelity of involutory
multi-qubit gates, including non-Clifford targets, for
which standard Clifford-based benchmarking does not apply.
%A.2
The method selects a compact set of experimentally
executable circuit measurements from a candidate pool
through offline training on a physically motivated
ensemble of noisy channels, and combines them into a
linear estimator with weights learned by ridge regression.
The training ensemble is an explicit and tunable component
of the protocol, incorporating prior knowledge about
dominant hardware noise mechanisms.
%A.3
The estimator is validated on \texttt{ibm\_kingston} using two Clifford validation benchmarks structurally related to the transpiled CCZ circuit, against independent Interleaved Randomized Benchmarking (IRB). Both show close
EFS--IRB agreement across a wide range of process infidelities, with an estimation precision of approximately $0.01$ over a process infidelity range of $0.02$--$0.20$.
EFS is subsequently applied directly to CCZ on the same device.
\end{abstract}

\maketitle

% ============================================================
% SECTIONS
% ============================================================
\input{sec_1}   % Introduction
\input{sec_2}   % Process Infidelity as Benchmarking Metric
\input{sec_3}   % Estimator Model
\input{sec_4}   % Candidate Feature Pool
\input{sec_5}   % Training Ensemble Generation
\input{sec_6}   % Feature Selection
\input{sec_7}   % Results
\input{sec_8_A}   % Conclusion
\input{sec_8_Ack} % Acknowledgements
% ============================================================
% BIBLIOGRAPHY
% ============================================================
\bibliographystyle{apsrev4-2}
\bibliography{ref_bib}
\input{sec_9_A} % Appendix A
\input{sec_9_B} % Appendix B
\input{sec_9_C} % Appendix C

\end{document}

%% file: sec_1.tex
% sec_1.tex — Introduction

\section{Introduction}
\label{sec:intro}

%1.1
Non-Clifford gates are essential for universal quantum
computation, yet their characterization remains a
significant challenge. Randomized benchmarking (RB) and
interleaved randomized benchmarking
(IRB)~\cite{magesan2011,magesan2012} are well-established
for Clifford gates, exploiting the algebraic structure of
the Clifford group to yield simple exponential decay models.
For non-Clifford targets this structure is absent, making
accurate and scalable characterization substantially more
difficult.
%1.2
Existing approaches span a broad range of resource
requirements and assumptions. Protected-gate benchmarking
in stabilizer codes~\cite{cross2016}, channel spectrum
benchmarking~\cite{gu2023}, and Pauli transfer character
benchmarking for involutory gates~\cite{ye2025ptcb} extend
benchmarking beyond the Clifford regime. Complete
characterization methods such as quantum process tomography
(QPT) and gate set tomography (GST) provide detailed
descriptions but scale exponentially with system size.

%1.3
The unfavorable scaling of complete methods has motivated
scalable alternatives. Machine-learning-assisted
frameworks~\cite{genois2021}, deep-reinforcement-learning
strategies for multiparameter estimation~\cite{cimini2024},
reduced-measurement learning for native CZ
gates~\cite{papic2023}, and featuremetric benchmarking
from coarse circuit descriptors~\cite{proctor2025}
demonstrate a growing interest in methods that trade a
moderate loss of information for improved scalability and
reduced experimental cost.

%1.4
Practical quantum computing frequently requires fast
feedback for calibration and error-correction workflows,
where the objective is not ultra-high-precision metrology.
While single- and two-qubit gates routinely achieve error
rates below $1\%$, three-qubit non-Clifford gates typically
exhibit error rates of several percent. In this regime, an
approximate but fast and scalable estimate of gate
performance is more valuable than a resource-intensive
characterization procedure.

%1.5
Feature selection is a central concept in statistics and
machine learning, where a large set of candidate observables
is reduced to a smaller yet informative subset for
prediction~\cite{guyon2003,bolon2019}. In this work we
propose Ensemble Feature Selection (EFS) as a practical
framework for estimating process infidelity of the non-Clifford CCZ gate, which
belongs to the family of involutory diagonal gates.
%1.5a
Although
cycle benchmarking (CB) has been demonstrated for
CCZ~\cite{hill2021ccphase,kim2022itoffoli}, its extension
beyond Clifford targets faces two difficulties: the required
correction-gate overhead can itself add
errors~\cite{hill2021ccphase, hashim2024qcvv}, and the
correction-operator construction is not fully specified in
the open
literature~\cite{erhard2019,wallman2016rc,wallman2019patent}.
These limitations motivate building an estimator for this
family of involutory targets, independent of which type the
target is. The estimator is then validated on the Clifford
members of this family, where IRB applies directly and does
not face the two CB difficulties described above.
 
%1.5b
The method selects a compact
subset of experimentally executable survival probability
measurements through offline training on a representative
ensemble of noisy channels, and combines them into a linear
estimator with weights learned by ridge regression subject
to an anchor constraint. A central aspect is that the
training ensemble is an explicit and tunable component of
the protocol, allowing prior knowledge about dominant
hardware noise mechanisms to be incorporated directly.
 
%1.6
The method is developed and demonstrated for diagonal
involutory gates and does not rely on Clifford properties
of the target. Validation is performed
on \texttt{ibm\_kingston} against independent IRB on two
Clifford circuits of comparable transpiled structure:
$\Iccz$~\cite{grigorova2026}, obtained from the transpiled CCZ circuit by
removing its virtual $R_z(\pm\pi/4)$ gates, and
CZ$(0,2)$, an embedded three-qubit Clifford gate.
%1.6a
The present validation exploits a particularly favorable
property of superconducting hardware. The only non-Clifford
ingredient distinguishing the transpiled CCZ circuit from
$\Iccz$ is a set of seven virtual $R_z(\pm\pi/4)$ rotations.
These rotations are implemented by classical frame updates
and therefore contribute negligible intrinsic error. Their
effect on the circuit is not to add noise but to redistribute
pre-existing coherent error components between the $X$ and
$Y$ sectors, while leaving diagonal $Z$-type errors
unchanged. Comparing EFS estimates for CCZ and $\Iccz$
therefore provides a controlled test of whether the
estimator remains sensitive to this redistribution, beyond
its IRB validation on Clifford targets.

%1.7
The paper is organized as follows.
Section~\ref{sec:metric} defines the process infidelity
and establishes its linear structure in the PTM
representation.
Section~\ref{sec:estimator} introduces the ideal and
practical estimator models.
Section~\ref{sec:pool} describes the candidate feature
pool and the anchor constraint.
Section~\ref{sec:ensemble} details the training ensemble.
Section~\ref{sec:selection} presents the feature-selection
and weight-learning procedure.
Section~\ref{sec:results} reports the experimental results.
Section~\ref{sec:conclusion} summarizes the findings and
outlines future directions.

%% file: sec_2.tex
% ============================================================
% sec_2.tex — Process Infidelity as the Benchmarking Metric
% ============================================================

\section{Process Infidelity as the Benchmarking Metric}
\label{sec:metric}

%2.1
We characterize gate performance via the process infidelity
\be
\label{eq:infidelity_def}
\eF(\cG,\Tgt)
=
1-\Fid(\cG,\Tgt),
\ee
where $\cG$ is the implemented noisy channel and $\Tgt$
is the ideal target gate, both represented as Pauli
transfer matrices (PTM).
%2.2
The process fidelity is
\be
\label{eq:fidelity_ptm}
\Fid(\cG,\Tgt)
=
\frac{1}{\Hd^2}
\trc\!\left(
\cG\,\Tgt^{-1}
\right),
\ee
where $\Hd = 2^n$ is the Hilbert-space dimension of
an $n$-qubit system.

%2.3
Writing the channel as $\cG = \chE \circ \Tgt$,
where $\chE$ is the error channel, gives
\be
\label{eq:infidelity_comp}
\eF(\cG,\Tgt) = 1 - \frac{1}{\Hd^2}\trc(\chE).
\ee

%2.4
Alternatively, in additive form $\cG = \Tgt + \Err$,
where $\Err$ is the error matrix,
\be
\label{eq:infidelity_additive}
\eF(\cG,\Tgt) = -\frac{1}{\Hd^2}\trc(\Err\,\Tgt^{-1}).
\ee

%2.5
Benchmarking protocols such as IRB report the average gate
infidelity $\varepsilon_{\mathrm{avg}}$, related~\cite{nielsen2002} to the process infidelity via
\be
\label{eq:nielsen}
\eF = \frac{d+1}{d}\,\varepsilon_{\mathrm{avg}}.
\ee

%2.6
The infidelity is thus linear in the PTM elements of $\cG$,
a property that motivates the estimator architecture of
Sec.~\ref{sec:estimator}.
%2.7
Both representations~\eqref{eq:infidelity_comp}
and~\eqref{eq:infidelity_additive}, together with the
conversion~\eqref{eq:nielsen}, are used interchangeably
throughout this work.

%% file: sec_3.tex
% sec_3.tex — Estimator Model

\section{Estimator Model}
\label{sec:estimator}

\subsection{Ideal Estimator}

%3.1
The ideal estimator assumes perfect state preparation,
perfect projective measurements, and exact tomographic
probabilities. In the Hilbert--Schmidt representation,
\be
\label{eq:prob_ideal}
\pobs_{jks}
=
\langle\langle \Pi_k | \cG | \rho_{js} \rangle\rangle,
\ee
where $j$ labels the prepared Pauli direction, $k$ the
measured projector, $s = \pm 1$ the preparation eigenvalue,
and $\{P_j\} = \{I,X,Y,Z\}^{\otimes n}$ the $n$-qubit
Pauli basis.

%3.2
The ideal preparation states and measurement projectors are
\be
\label{eq:prep_meas}
\rho_{js} = \frac{1}{\Hd}(I + s P_j),
\qquad
\Pi_k = \frac{I + P_k}{2}.
\ee

%3.3
The PTM elements are recovered from the difference of the two
preparation signs,
\be
\label{eq:ptm_elements}
\cG_{kj} = \pobs_{jk,+} - \pobs_{jk,-}.
\ee

%3.4
Substituting into Eq.~\eqref{eq:fidelity_ptm} gives
\be
\label{eq:infidelity_sum}
\eF(\cG,\Tgt)
=
1 - \frac{1}{\Hd^2}
\sum_{j,k} \cG_{kj}\,(\Tgt^{-1})_{jk}.
\ee

%3.5
Only triples $(j,k,s)$ for which $(\Tgt^{-1})_{jk} \neq 0$
contribute; relabeling these as $\midx \equiv (j,k,s)$
and defining $\pobs_\midx = \pobs_{jks}$, the infidelity takes the
exact linear form
\be
\label{eq:ideal_estimator}
\eF
=
\wzero + \sum_{\midx=1}^{\Mid} \wfeat_\midx\, \pobs_\midx,
\ee
with offset and weights determined analytically by the target,
\be
\label{eq:ideal_weights}
\wzero = 1 - \frac{1}{\Hd^2},
\qquad
\wfeat_\midx = -\frac{1}{\Hd^2}\,\Tgt_{jk}\,(-1)^\midx.
\ee

%3.6
For a three-qubit Clifford gate the PTM is a signed permutation
matrix, requiring $\Mid=126$ feature circuits; the non-Clifford
CCZ gate has a dense PTM with 231 nontrivial elements, requiring
$\Mid=462$.
%3.7
This rapid growth motivates the practical estimator introduced
below, which replaces the full tomographic feature set by a
compact experimentally selected subset.

\subsection{Practical Estimator}
\label{sec:practical}

%3.8
In practice, the ideal tomographic measurements are not directly
realizable; we retain the linear structure and write
\be
\label{eq:practical_estimator}
\widehat{\eF}
= \wzero + \sum_{\midx=1}^{\Mid} \wfeat_\midx\, \pobs_\midx(\cG),
\ee
where $\pobs_\midx(\cG)$ is a probability from a circuit including
the noisy gate and $\Mid$ is the number of selected features.

%3.9
In contrast to the ideal estimator, the weights $\wfeat_\midx$,
offset $\wzero$, and feature indices are no longer fixed analytically
--- they are optimized jointly on a physically motivated training
ensemble, described in Sec.~\ref{sec:ensemble}.

%3.10
Since $\eF(\cG,\Tgt) = \eF(\chE,I)$, the quantity of
interest depends only on the error channel $\chE$, not
on $\Tgt$ itself. We therefore impose the anchor condition
\be
\label{eq:anchor_condition}
\widehat{\eF}(\Tgt) = \widehat{\eF}(I) = 0,
\ee
and define
\be
\label{eq:delta_p}
\Delta\pobs_\midx = \pobs_\midx(\cG) - \pobs_\midx(I).
\ee

%3.11
The anchored estimator then takes the form
\be
\label{eq:anchored_estimator}
\widehat{\eF}
= \sum_{\midx=1}^{\Mid} \wfeat_\midx\, \Delta\pobs_\midx,
\ee
where $\wzero$ drops out.

%% file: sec_4.tex
% sec_4.tex — Candidate Feature Pool

\section{Candidate Feature Pool}
\label{sec:pool}

%4.1
Each candidate feature $\midx$ corresponds to a survival probability
\be
\label{eq:survival_prob}
\pobs_\midx(\cG)
=
\langle\langle \Pi_s | (\cG\,\chZ_\midx)^{\Lrep} | \rho_s \rangle\rangle,
\ee
where $\rho_s$ is the prepared state, $\Pi_s$ the measurement
projector, $\chZ = \bigotimes_q R_z(\phi_q)$ a virtual $Z$ mixer,
and $\Lrep$ the repetition depth.
%4.2
Virtual $Z$ rotations are particularly convenient on IBM
superconducting devices, where they are implemented as frame
updates and are effectively ideal~\cite{mckay2017}.

%4.2a
Beyond this practical convenience, the virtual $\chZ_\midx$
operations serve as \emph{analysis mixers}. By rotating
coherent $X/Y$ error components while leaving diagonal
$Z$-type errors invariant, they generate multiple
experimentally accessible projections of the same underlying
coherent error structure. This enriches the feature pool
without introducing additional physical noise.

%4.2b
This role is closely related to the Pauli conjugation induced
by the virtual $R_z(\pm\pi/4)$ rotations appearing in the
transpiled CCZ circuit (Sec.~\ref{sec:results}). Although the
two sets of rotations act at different locations in the
circuit and use different angle sets, both belong to the same
family of Pauli-plane rotations: they leave the $Z$ axis
invariant and rotate only the $X$--$Y$ subspace. This
algebraic correspondence provides a natural explanation for
why the selected feature family is sensitive to the
redistribution of coherent error components introduced by the
non-Clifford implementation.

%4.3
Preparation states and mixer angles are restricted to the
discrete sets
\be
\label{eq:prep_basis}
\{|0\rangle, |1\rangle, |+\rangle, |i\rangle\}^{\otimes n},
\qquad
\chZ \in \{I,\, S,\, Z,\, S^\dagger\}^{\otimes n},
\ee
keeping the candidate pool of manageable size.

%4.4
Both targets commute with $\chZ$, so the anchor condition
\be
\label{eq:anchor_circuit}
(\Tgt\,\chZ_m)^{\Lrep} = \chZ_m^{\Lrep} = I
\ee
is satisfied for $\Lrep \in \{4, 8, 12, \ldots\}$.
%4.4а
The unitary matrices of
the two targets in the computational basis are
\be
\label{eq:cz02_matrix}
\mathrm{CZ}(0,2) = \mathrm{diag}(1, 1, 1, 1, 1, -1, 1, -1),
\ee
\be
\label{eq:ccz_matrix}
\mathrm{CCZ}(0,1,2) = \mathrm{diag}(1, 1, 1, 1, 1, 1, 1, -1),
\ee
and their PTM representations $\Tgt$ are used throughout.

%4.5
Since $\eF(\cG,\Tgt) = \eF(\chE,I)$ for any target
satisfying the anchor condition $(\Tgt\,\chZ_m)^4=I$,
the process infidelity is directly accessible, and the
estimator as a whole does not distinguish $\Tgt$ from $I$.
%4.6
With $\Lrep \in \{4, 8\}$ and $n=3$ qubits, the filtered pool
contains 8192 candidate feature circuits.

%4.7
Rather than a full Pauli twirl as in cycle
benchmarking~\cite{erhard2019}, we compensate for the
incomplete averaging through four complementary elements:
discrete $Z$ rotations from $\{I, S, Z, S^\dagger\}^{\otimes n}$,
a hardware-informed training ensemble
(Sec.~\ref{sec:ensemble}), greedy selection of
shot-noise-aware informative features
(Sec.~\ref{sec:selection}), and learned weights.

%4.8
The choice of $\{0, \pm\pi/2, \pi\}$ for the mixer angles
is not claimed to be optimal; a coarser discretization ---
for instance $120^\circ$ rotations $\{0, 2\pi/3, 4\pi/3\}$
with $\Lrep \in \{3, 6, 9, \ldots\}$ --- would reduce the
pool size and is left for future investigation.

%4.9
Measuring $\pobs_\midx(I)$ alongside $\pobs_\midx(\cG)$
mitigates static SPAM and readout errors to first order,
provided these errors are shared by the two circuits.
Writing $\rho \to \rho + \delta\rho$ and
$\Pi \to \Pi + \delta\Pi$,
\be
\label{eq:spam_expansion}
\Delta p = \trc[\Pi\, \delta\cG\, \rho]
+ \trc[\delta\Pi\, \delta\cG\, \rho]
+ \trc[\Pi\, \delta\cG\, \delta\rho]
+ O(\delta^3),
\ee
where $\delta\cG = (\cG\,\chZ)^{\Lrep} - I$, so preparation
and readout errors enter only at second order.

%% file: sec_5.tex
% ============================================================
% sec_5.tex — Training Ensemble Generation
% ============================================================

\section{Training Ensemble Generation}
\label{sec:ensemble}

%5.1
The training ensemble serves as a hardware-informed prior over
the space of noise channels associated with the target gate.
It should cover the physically relevant coherent and dissipative
error directions without including unrealistic noise, which
would dilute training accuracy.

% ============================================================
\subsection{Channel Construction}
\label{sec:ensemble_general}

%5.2
Each ensemble member is a noisy channel constructed in
compositional form as
\be
\label{eq:channel_compositional}
\cG = \chEu \circ \chEv \circ \Tgt,
\ee
where $\chEu$ and $\chEv$ are the coherent and dissipative
error channels, respectively.

% ============================================================
\subsection{Dissipative Noise Model}
\label{sec:ensemble_dissip}

%5.3
The dissipative component is modeled as a tensor-product
amplitude-damping and dephasing channel acting independently
on each qubit,
\be
\label{eq:dissip_channel}
\chEv
=
\chEv^{(0)}
\otimes
\chEv^{(1)}
\otimes
\chEv^{(2)},
\ee
with each single-qubit factor parametrized by $T_1^{(q)}$
and $T_2^{(q)} \le 2T_1^{(q)}$, sampled from distributions
consistent with the reported hardware values~\cite{IBMQuantum,krantz2019}.
%5.4
Additional stochastic Pauli channels may be incorporated
when supported by the calibration data.

% ============================================================
\subsection{Coherent Noise Model}
\label{sec:ensemble_unitary}

%5.5
The coherent component is a unitary error
\be
\label{eq:unitary_err}
U_{\mathrm{err}} = e^{-i\theta \tilde{H}},
\ee
where $\tilde{H} = H/\|H\|_F$ is the normalized Hamiltonian
and $\theta \in [0, \theta_{\mathrm{max}}]$ controls the
coherent error strength.
%5.6
The Hamiltonian is expanded in the Pauli basis as
\be
\label{eq:hamiltonian}
H = \sum_j \theta_j P_j,
\ee
with coefficients drawn as
\be
\label{eq:theta_sigma}
\theta_j \sim \mathcal{N}(0, \sigma_g^2)
\ee
for $j$ in group $g$, defining the \emph{Pauli spectral
structure} of the ensemble , sampled over $K=5000$ noisy channel realizations $\{\cG_k\}_{k=1}^K$.
%5.7
The PTM of $\chEu$ is obtained from $U_{\mathrm{err}}$ as
\be
\label{eq:ptm_unitary}
(\chEu)_{ij}
=
\frac{1}{d}
\trc\!\left(
P_i\, U_{\mathrm{err}}\, P_j\, U_{\mathrm{err}}^\dagger
\right).
\ee

%5.8
The variance scales $\sigma_g$ used in this work are listed in
Table~\ref{tab:pauli_scales}. Conditional $Z$ rotations are
assigned the largest scale, motivated by the known dominance
of conditional phase errors in controlled-phase
gates~\cite{krinner2020,krantz2019}.

\begin{table}[htbp]
\caption{\justifying Pauli spectral structure of the noise Hamiltonian.
Conditional $Z$ rotations are assigned the largest variance scale,
motivated by the dominance of conditional phase errors in
controlled-phase gates~\cite{krinner2020,krantz2019}.
This choice is not claimed to be optimal.}
\label{tab:pauli_scales}
\begin{ruledtabular}
\begin{tabular}{lll}
Pauli sector & Operators & $\sigma_g$ \\
\hline
Conditional $Z$ & ZZI, ZIZ, IZZ, ZZZ & 1.00 \\
Local $Z$        & ZII, IZI, IIZ       & 0.25 \\
Single-qubit $XY$ & XII, IXI, IIX, YII, IYI, IIY & 0.20 \\
Two-qubit $XY$   & XXI, XIX, IXX, YYI, \ldots   & 0.12 \\
Mixed            & XZI, ZXI, XIZ, \ldots         & 0.08 \\
Non-diagonal 3Q  & XXX, XYZ, \ldots              & 0.05 \\
\end{tabular}
\end{ruledtabular}
\end{table}

% ============================================================
\subsection{Involutory Error Structure}
\label{sec:ensemble_involutory}

%5.9
Since the targets considered in this work are involutory
($\Tgt^2 = I$), it is convenient to present the generated
ensemble in a compact four-parameter view.

%5.10
The additive error $\Err = \cG - \Tgt$ decomposes into
anticommuting and commuting parts,
\be
\Err = \Ea + \Ec,
\ee
where
\be
\label{eq:Ea_def}
\Ea = \frac{1}{2}\left(\Err - \Tgt\,\Err\,\Tgt\right),
\qquad
\Ec = \frac{1}{2}\left(\Err + \Tgt\,\Err\,\Tgt\right).
\ee

%5.11
Since $\trc(\Ea\,\Tgt) = 0$, only the commuting part
contributes to the process infidelity,
\be
\label{eq:eps_ac}
\eF = -\frac{1}{\Hd^2}\trc(\Ec\,\Tgt).
\ee

%5.12
The decomposition is summarized by two scalar pairs.
The anticommuting and commuting scalars are
\be
\label{eq:a_def}
\acom := \frac{\trc(\Ea\,\Ea^{\tp})}{2\Hd^2},
\qquad
\com := -\frac{\trc(\Err\,\Tgt)}{\Hd^2} - \frac{\|\Ea\|_F^2}{2\Hd^2},
\ee
so that $\eF = \acom + \com$.

%5.13
Independently, from the unitarity condition $\cU\cU^{\tp}=I$
with $\cU = \Tgt + \Eu$, the coherent and incoherent scalars are
\be
\label{eq:u_def}
\uni := \frac{\trc(\Eu\Eu^{\tp})}{2\Hd^2},
\qquad
\nuni := -\frac{\trc(\Err\Tgt)}{\Hd^2} - \frac{\|\Eu\|_F^2}{2\Hd^2},
\ee
so that $\eF = \uni + \nuni$.

\begin{figure}[htbp]
\centering
\includegraphics[width=\columnwidth]{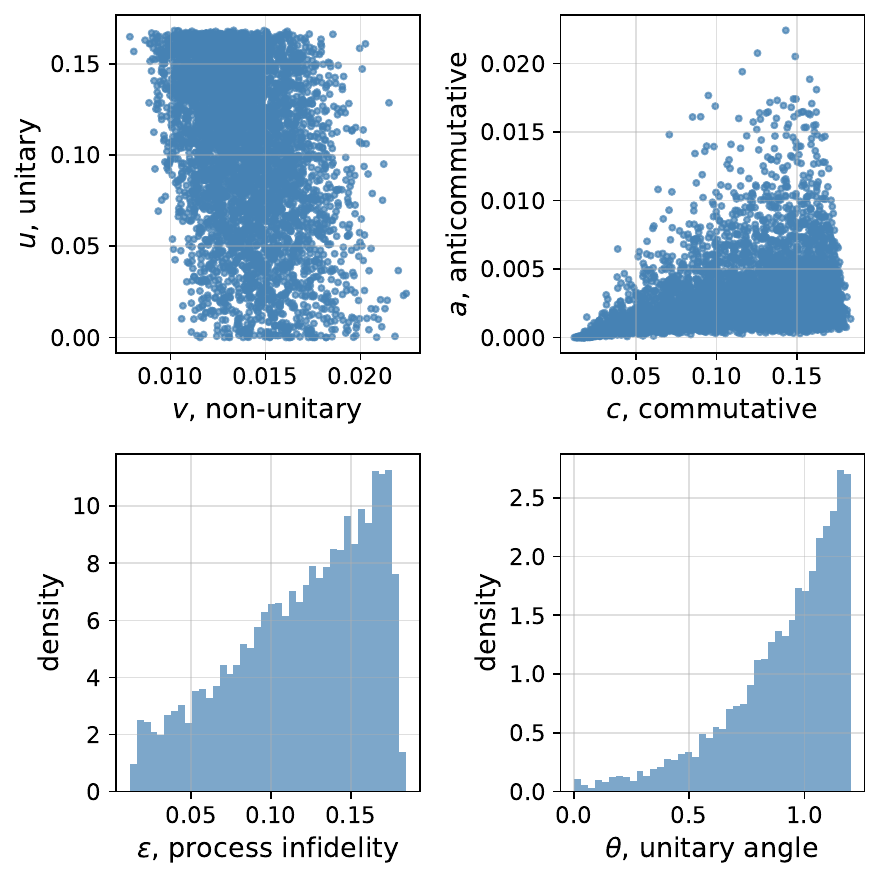}
\caption{\justifying Error decomposition of the training ensemble ($K=5000$ realizations).
Top left: $(u,v)$ scatter separating coherent ($u$) and incoherent
($v$) contributions. Top right: $(a,c)$ scatter separating
anticommuting ($a$) and commuting ($c$) parts relative to the
involutory target. Bottom left: distribution of the sampled
infidelity $\varepsilon$, biased toward larger errors to improve
training coverage. Bottom right: distribution of the coherent
error angle $\theta$.}
\label{fig:uvca_eps_theta}
\end{figure}

%5.14
The pair $(\uni,\nuni)$ separates coherent and incoherent
contributions; $(\acom,\com)$ separates anticommuting and
commuting parts relative to the involutory target structure.
Together, the four scalars $(\uni,\nuni,\acom,\com)$ provide
a compact diagnostic of the generated ensemble, visualized in
Fig.~\ref{fig:uvca_eps_theta}.

%5.15
The upper bound is set to $\varepsilon_{\mathrm{max}} \approx 0.2$,
consistent with the typical error range of present-day 3Q
operations. Ensemble members are sampled more densely at larger
infidelities, since the noise-space volume grows with distance
from the ideal gate.

%% file: sec_6.tex
\section{Feature Selection and Weight Learning}
\label{sec:selection}

%6.1
Given the feature pool of Sec.~\ref{sec:pool} and the training
ensemble $\{\cG_\ki\}_{\ki=1}^{\Ke}$ of
Sec.~\ref{sec:ensemble}, the task is to select $\Mid$ features
and learn their weights such that
Eq.~\eqref{eq:anchored_estimator} accurately predicts
$\varepsilon$ on unseen data.
%6.2
The procedure combines ridge regression for weight learning
with greedy forward selection for the feature subset, using
a shot-noise-aware RMSE criterion that penalizes features
sensitive to shot noise.

\subsection{Weight Learning}
\label{sec:weights}

%6.4
For a fixed subset of $\Mid$ features, the weights are
obtained by ridge regression~\cite{tikhonov1977,hoerl1970ridge},
\be
\label{eq:ridge}
\mathbf{w}
=
\left(\mathbf{P}^{\mathsf{T}}\mathbf{P} + \lambda I\right)^{-1}
\mathbf{P}^{\mathsf{T}} \boldsymbol{\varepsilon},
\ee
where $\mathbf{P}$ is the $\Ke \times \Mid$ matrix of
$\Delta\pobs_\midx^{(\ki)}$ values, $\boldsymbol{\varepsilon}$
is the vector of ensemble infidelities, and $\lambda$ prevents
instability when features are correlated.

\subsection{Greedy Feature Selection}
\label{sec:greedy}

%6.5
Exhaustive search over 8192 candidates is computationally
intractable; we use greedy forward
selection~\cite{kohavi1997wrappers,guyon2003}, adding at each
step the candidate that minimizes the shot-noise-aware RMSE,
\be
\label{eq:rmse_criterion}
\begin{aligned}
\mathrm{RMSE}
=&\\[+1 mm]
&\hspace{-1.5cm}
\sqrt{
\frac{1}{\Ke}\sum_{\ki=1}^{\Ke}
\left(\varepsilon^{(\ki)}-\widehat{\varepsilon}^{(\ki)}\right)^2
+
\sum_{\midx=1}^{\Mid}
\wfeat_\midx^2
\frac{\sigma_\midx^2(\cG)+\sigma_\midx^2(I)}{\Ns}
}
\end{aligned}
\ee
where $\sigma_\midx^2 = \pobs_\midx(1-\pobs_\midx)$ is the
Bernoulli variance and $\Ns$ the number of shots per circuit;
since $\Delta\pobs_\midx$ is the difference of two independent
measurements, their variances add.

\begin{figure}[t]
\centering
\includegraphics[width=0.48\textwidth]{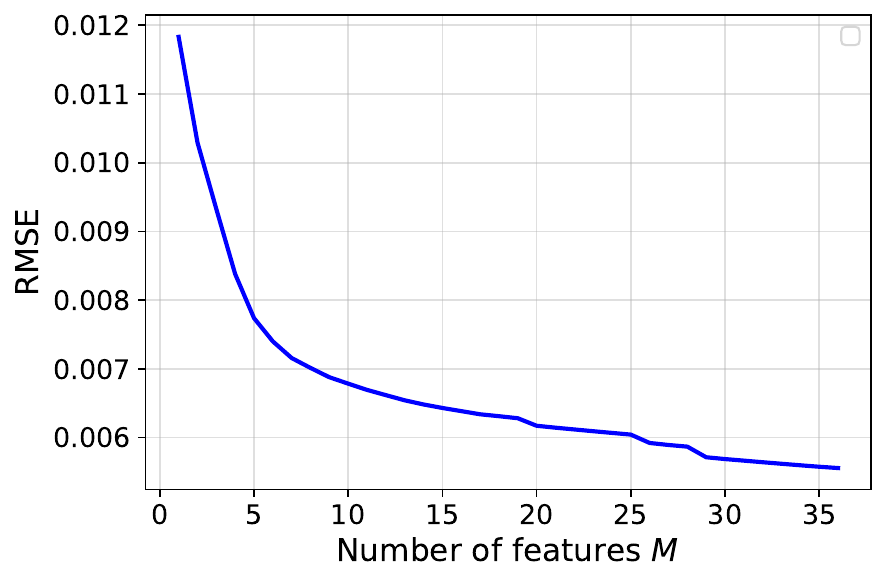}
\caption{\justifying Combined RMSE as a function of the number of
selected features $\Mid$ for $\Lrep \in {4,8}$. The
training ensemble contains noisy realizations of
$I$, CZ$(0,2)$ and CCZ generated with different random
seeds. Consequently, a single EFS estimator is learned
simultaneously for all three targets, resulting in the
same RMSE curve.}
\label{fig:rmse_vs_m}
\end{figure}

\begin{table}[t]
\caption{
Selected EFS features for the CZ$(0,2)$ target with $\Mid=24$.
}
\label{tab:efs_features}

\centering
\small
\setlength{\tabcolsep}{4pt}
\renewcommand{\arraystretch}{1.05}

\begin{tabular}{c c c c c}
\hline
$m$ & $\psi_m, \rho_m$ & $\phi_m\{q_2,q_1,q_0\}$ & $L_m$ & $w_m \times 10^2$ \\
\hline

1  & \(|+\rangle|i\rangle|i\rangle\)
   & \((-\pi/2,0,0)\)
   & 4
   & \(-3.53\) \\

2  & \(|1\rangle|0\rangle|0\rangle\)
   & \((0,0,\pi)\)
   & 4
   & \(-2.34\) \\

3  & \(|i\rangle|+\rangle|i\rangle\)
   & \((+\pi/2,0,-\pi/2)\)
   & 4
   & \(-4.48\) \\

4  & \(|+\rangle|+\rangle|+\rangle\)
   & \((+\pi/2,\pi,-\pi/2)\)
   & 8
   & \(+3.32\) \\

5  & \(|1\rangle|1\rangle|1\rangle\)
   & \((\pi,0,0)\)
   & 4
   & \(-2.20\) \\

6  & \(|+\rangle|+\rangle|i\rangle\)
   & \((0,+\pi/2,\pi)\)
   & 4
   & \(-3.28\) \\

7  & \(|0\rangle|1\rangle|1\rangle\)
   & \((0,-\pi/2,0)\)
   & 4
   & \(-1.67\) \\

8  & \(|+\rangle|+\rangle|+\rangle\)
   & \((\pi,+\pi/2,-\pi/2)\)
   & 4
   & \(-2.68\) \\

9  & \(|1\rangle|1\rangle|0\rangle\)
   & \((\pi,+\pi/2,-\pi/2)\)
   & 4
   & \(-3.47\) \\

10 & \(|+\rangle|i\rangle|+\rangle\)
   & \((+\pi/2,-\pi/2,\pi)\)
   & 4
   & \(-3.60\) \\

11 & \(|0\rangle|0\rangle|0\rangle\)
   & \((+\pi/2,+\pi/2,\pi)\)
   & 8
   & \(-9.92\) \\

12 & \(|0\rangle|1\rangle|1\rangle\)
   & \((+\pi/2,0,-\pi/2)\)
   & 4
   & \(-2.92\) \\

13 & \(|i\rangle|i\rangle|+\rangle\)
   & \((-\pi/2,\pi,+\pi/2)\)
   & 4
   & \(-1.81\) \\

14 & \(|1\rangle|1\rangle|0\rangle\)
   & \((0,-\pi/2,\pi)\)
   & 4
   & \(-2.08\) \\

15 & \(|0\rangle|0\rangle|1\rangle\)
   & \((0,-\pi/2,0)\)
   & 4
   & \(-1.88\) \\

16 & \(|0\rangle|i\rangle|i\rangle\)
   & \((0,0,0)\)
   & 4
   & \(+1.78\) \\

17 & \(|1\rangle|i\rangle|i\rangle\)
   & \((-\pi/2,0,\pi)\)
   & 4
   & \(+1.61\) \\

18 & \(|i\rangle|+\rangle|+\rangle\)
   & \((0,\pi,+\pi/2)\)
   & 4
   & \(-3.10\) \\

19 & \(|i\rangle|0\rangle|+\rangle\)
   & \((0,0,\pi)\)
   & 4
   & \(+1.67\) \\

20 & \(|i\rangle|1\rangle|+\rangle\)
   & \((0,-\pi/2,\pi)\)
   & 4
   & \(+1.52\) \\

21 & \(|i\rangle|+\rangle|0\rangle\)
   & \((+\pi/2,0,\pi)\)
   & 4
   & \(+1.76\) \\

22 & \(|+\rangle|+\rangle|1\rangle\)
   & \((+\pi/2,0,\pi)\)
   & 4
   & \(+1.74\) \\

23 & \(|i\rangle|i\rangle|i\rangle\)
   & \((-\pi/2,0,+\pi/2)\)
   & 4
   & \(-1.85\) \\

24 & \(|1\rangle|+\rangle|i\rangle\)
   & \((\pi,0,\pi)\)
   & 8
   & \(-0.45\) \\

\hline
\end{tabular}
\end{table}

%6.6
Since every target considered in this work satisfies
Eq.~\eqref{eq:anchor_circuit}, the estimator as a whole does
not distinguish a target gate $\Tgt$ from $I$.
%6.7
Moreover, the estimator is neither Clifford- nor non-Clifford-specific: since it does not distinguish $\Tgt$ from $I$ (item above), its accuracy can be assessed on any target satisfying the anchor condition. We therefore use Clifford circuits that are structurally related to the
transpiled CCZ implementation to validate the estimator against independent IRB measurements.

%6.8
The validation benchmarks we use are $\Iccz$ and CZ$(0,2)$.
$\Iccz$, obtained from the transpiled CCZ circuit by removing its virtual $R_z(\pm\pi/4)$ gates, and CZ$(0,2)$, an embedded three-qubit Clifford gate, are both implemented on the same device and from the same native gate set as CCZ; the circuits are shown in \ref{sec:appendix_circ}.

%6.9
The combined RMSE as a function of $\Mid$ is shown in Fig.~\ref{fig:rmse_vs_m} for $\Lrep \in \{4,8\}$. The training ensemble combines noisy realizations of $I$, CZ$(0,2)$ and CCZ generated with different random
seeds. Feature selection and weight optimization are performed jointly on the combined dataset, resulting in
a common RMSE curve.
%6.10
The selected features and their weights for $\Mid = 24$ are
listed in Table~\ref{tab:efs_features}.
%6.11
We use $\Mid=24$ throughout, as increasing to $\Mid=36$
yields no substantial improvement
(Table~\ref{tab:m_selection}). For completeness, the
scatter plot and feature table for $\Mid=36$ are provided
in Appendix~\ref{sec:appendix_b}.

% ------------------------------------------------------------
\subsection{Pre-experimental Simulation}
\label{sec:preexp_sim}
 
\begin{figure}[t]
\centering
\includegraphics[width=0.48\textwidth]{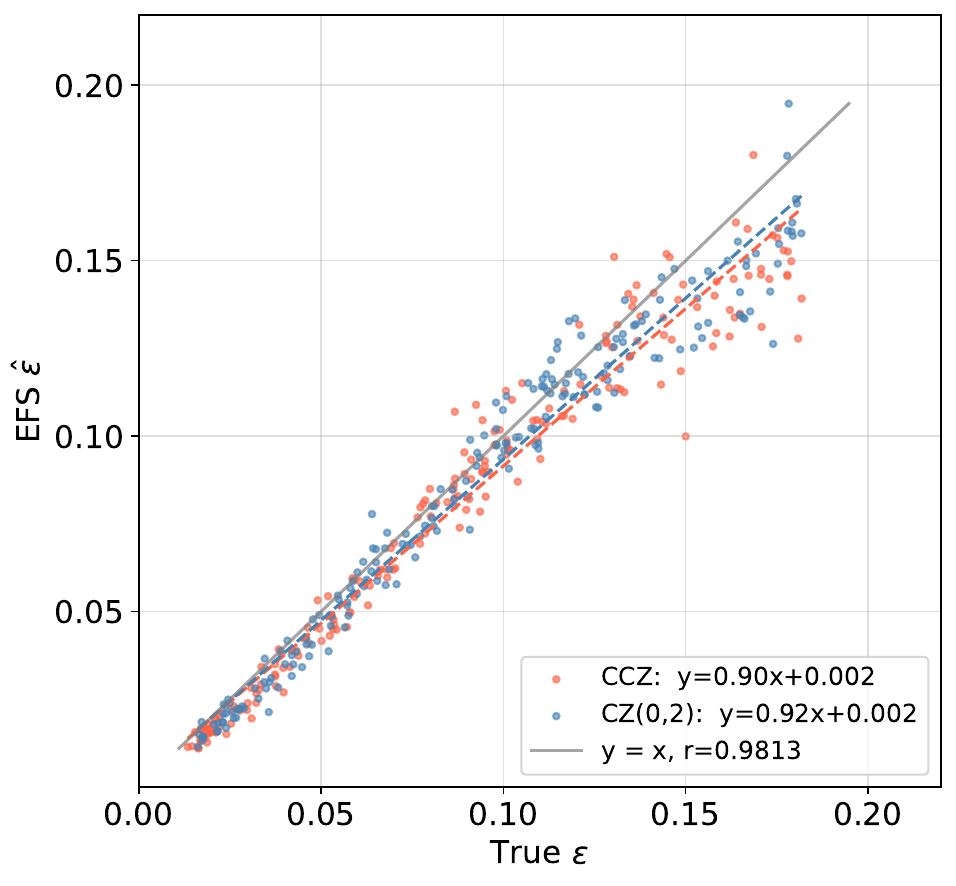}
\caption{\justifying EFS estimates vs true process infidelity
for CCZ (red) and CZ$(0,2)$ (blue) from Aer Simulator with
independent test ensemble and SPAM/readout errors.
Both targets yield slope $\approx 0.9$--$0.92$, confirming
that the estimator is equally applicable to non-Clifford gates.}
\label{fig:ccz_cz_aer}
\end{figure}

%6.12
Before proceeding to hardware experiments, the estimator
is validated via Aer Simulator~\cite{qiskit_aer} using
a holdout test ensemble~\cite{kohavi1995} with an
independent random seed. The simulator generates
realistic noise channel realizations with randomly
sampled SPAM, readout, and shot-noise errors in
hardware-typical ranges, and the true process infidelity
$\varepsilon$ is exactly known, serving as ground truth.
%6.13
Fig.~\ref{fig:ccz_cz_aer} shows EFS estimates versus
true $\varepsilon$ for both CCZ and CZ$(0,2)$. The two
targets yield nearly identical results: slope $0.9$ and
$0.92$, respectively, with a common offset of $0.002$,
confirming that the estimator applies consistently
across both targets prior to hardware deployment.

%% file: sec_7.tex
% ============================================================
% sec_7.tex — Results
% ============================================================

\section{Results}
\label{sec:results}

% ------------------------------------------------------------
\subsection{Experimental Setup}
\label{sec:setup}

%7.1
Experiments are performed on \texttt{ibm\_kingston} ---
a 156-qubit superconducting processor of type
Heron r2~\cite{IBMQuantum}.

%7.2
EFS measurements are performed with $\Ns = 2000$ shots per
circuit. IRB is performed at depths $\{1, 2, 4, 8, 16, 32, 64\}$, each with 15 random Clifford sequences and $\Ns = 1000$per circuit, immediately after the EFS session to minimize drift; the resulting $\varepsilon_{\mathrm{avg}}$ from IRB
is converted to process infidelity via Eq.~\eqref{eq:nielsen}.

%7.3
The EFS estimator is compared against independent IRB
measurements for the validation benchmarks
$\Iccz$ and CZ$(0,2)$ introduced in
Sec.~\ref{sec:selection}; their transpiled circuits are
shown in Appendix~\ref{sec:appendix_circ}.
%7.3a
All hardware results use $\Mid=24$
(Table~\ref{tab:m_selection}); the scatter plot and
feature table for $\Mid=36$ are given in
Appendix~\ref{sec:appendix_b}.

\begin{table}[t]
\centering
\caption{\justifying EFS--IRB performance on \texttt{ibm\_kingston}
for CZ$(0,2)$ as a function of $\Mid$, providing empirical
justification for the choice $\Mid=24$. The $\Mid=36$
scatter plot and feature table are given in
Appendix~\ref{sec:appendix_b}.}
\label{tab:m_selection}
\begin{tabular}{cccc}
\hline
$\Mid$ & Slope $\alpha$ & $r$ & $\mathrm{RMSE}_{\mathrm{cal}}$ \\
\hline
6  & 0.424 & 0.804 & 0.0235 \\
24 & 0.928 & 0.914 & 0.0081 \\
36 & 1.008 & 0.930 & 0.0078 \\
\hline
\end{tabular}
\end{table}

%7.4
EFS and IRB are performed on all valid triples
$(q_0, q_1, q_2)$ where $q_0$--$q_1$ and $q_1$--$q_2$ are
connected edges, excluding qubits and edges flagged as faulty.
\texttt{ibm\_kingston} yields 204 valid triples,
grouped into disjoint sets; for each group,
EFS($\Iccz$), IRB($\Iccz$), and EFS(CCZ) are executed
sequentially, ensuring near-identical hardware conditions
within each group. For CZ$(0,2)$, only EFS and IRB are
performed, without the additional EFS(CCZ) step.

%7.4a
To quantify the agreement between EFS and IRB on hardware,
we use the calibrated RMSE
\be
\label{eq:rmse_cal}
\mathrm{RMSE}_{\mathrm{cal}}
=
\frac{\sigma_{\mathrm{EFS}}}{\alpha}\sqrt{1-r^2},
\ee
where $\sigma_{\mathrm{EFS}}$ is the standard deviation of
the EFS estimates, $\alpha$ the slope, and $r$ the Pearson
correlation of the EFS--IRB scatter.

% ------------------------------------------------------------
\subsection{EFS--IRB Validation on $\Iccz$ and CZ$(0,2)$}
\label{sec:iccz}

%7.5
The closest noisy variant with a Clifford target that we
have identified is $\Iccz$
(Sec.~\ref{sec:selection}, Appendix~\ref{sec:appendix_circ}).
Removing the seven virtual $R_z(\pm\pi/4)$ gates from the
transpiled CCZ circuit yields $\Iccz$ while preserving the
same native-gate implementation and all associated noise
sources. Since these virtual rotations contribute negligible
intrinsic error (Sec.~\ref{sec:intro}), the dominant physical
difference between CCZ and $\Iccz$ is not additional noise
but a redistribution of pre-existing $X/Y$ error components
(Fig.~\ref{fig:ccz_trans_draw}). Comparing the two circuits
therefore isolates the estimator's response to this
redistribution.

%7.6
As a complementary check, CZ$(0,2)$
(Appendix~\ref{sec:appendix_circ}) is used to extend the
validation to a lower range of $\varepsilon$, not well
covered by $\Iccz$.

\begin{figure}[t]
\centering
\includegraphics[width=0.48\textwidth]{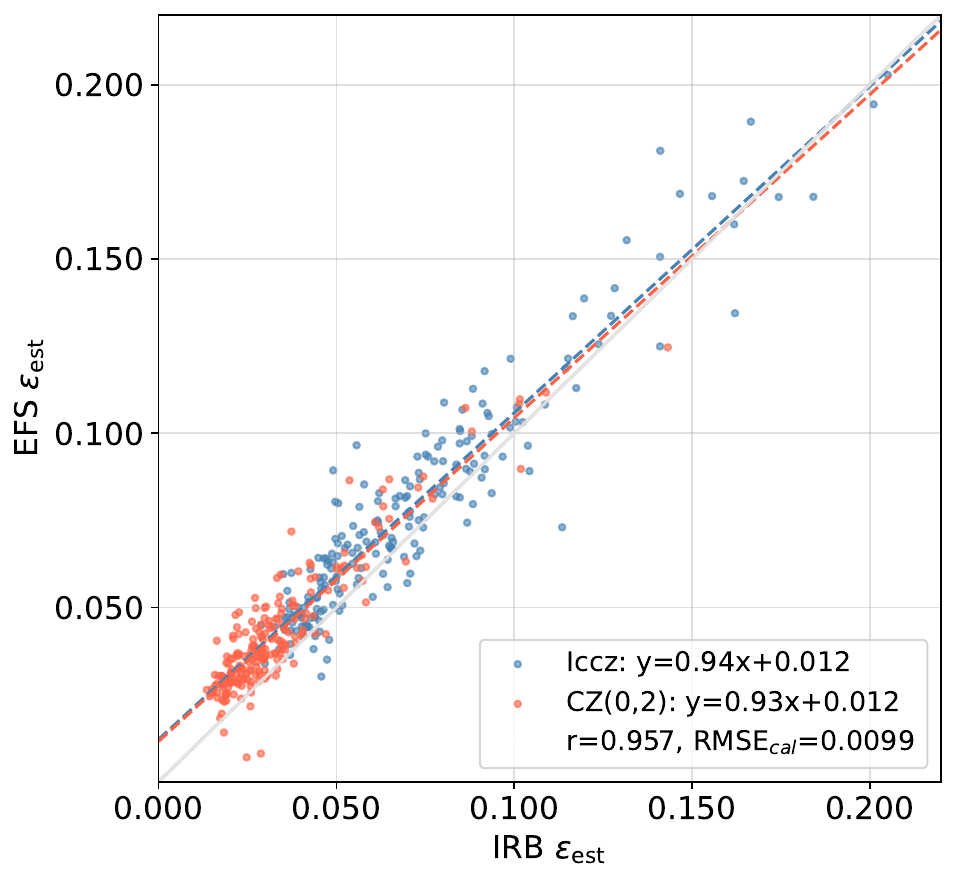}
\caption{\justifying EFS vs IRB for $\Iccz$ and CZ$(0,2)$ on
\texttt{ibm\_kingston} with $\Mid=24$. Fitted slopes and
intercepts are very similar for the two targets
($\alpha=0.94$ and $\alpha=0.93$, both with intercept $0.012$);
combined $r=0.957$, $\mathrm{RMSE}_{\mathrm{cal}}=0.0099$,
over $\varepsilon\approx 0.02$--$0.20$.}
\label{fig:iccz_cz02_kingston}
\end{figure}

%7.7
Fig.~\ref{fig:iccz_cz02_kingston} shows the combined
EFS--IRB comparison for $\Iccz$ and CZ$(0,2)$ on
\texttt{ibm\_kingston} with $\Mid=24$. The fitted slopes and intercepts for the two targets are $\alpha=0.94$ and $\alpha=0.93$, both with intercept $0.012$, and the combined data give Pearson correlation $r=0.957$ and $\mathrm{RMSE}_{\mathrm{cal}}=0.0099$ over
$\varepsilon\approx 0.01$--$0.20$. The combined EFS--IRB agreement across both benchmarks provides the primary validation result, with an estimation precision of approximately $0.01$ over $\varepsilon\approx 0.02$--$0.20$.

%7.8
Neither IRB nor EFS on $\Iccz$ measures CCZ directly;
the observed agreement instead provides an empirical
estimate of the accuracy expected when applying EFS to
the full CCZ circuit, although not necessarily at the
same infidelity value.

% ------------------------------------------------------------

\subsection{Clifford-to-Non-Clifford Transfer Validation}
\label{sec:transfer}

%7.9
To understand the Clifford-to-non-Clifford accuracy transfer in EFS, we first examine the corresponding argument in Cycle Benchmarking (CB).

%7.10
In CB, the Clifford and non-Clifford protocols are structurally identical, except for the final correction (nulling) gate that restores the ideal circuit to the identity. For non-Clifford targets, this correction may require a substantially longer physical implementation than in the Clifford case, thereby introducing additional noise~\cite{hill2021ccphase,hashim2024qcvv}.
%7.11
Despite this additional source of noise, the accuracy established for the Clifford case is accepted to carry over to the non-Clifford regime.

%7.12
EFS follows the same CB transfer principle: both the Clifford and non-Clifford protocols form ideal identity circuits by construction, Eq.~\eqref{eq:anchor_circuit}, without the additional correction gate required by CB. If CB's transfer of accuracy is accepted despite the noise introduced by that gate, the same transfer is even more naturally justified for EFS, where no additional noise accompanies the Clifford-to-non-Clifford transition.
%7.13
Applying the same validation criterion for Clifford-to-non-Clifford transfer to both CB and EFS, it follows that the accuracy established for EFS on Clifford targets carries over directly to the non-Clifford CCZ target.

%7.14
Our claim is therefore not to compare the accuracy of EFS and CB, but to show that the Clifford-to-non-Clifford transfer in EFS follows the same transfer argument as CB.

\subsection{EFS Estimation for CCZ}
\label{sec:ccz_hardware}

\begin{figure}[t]
\centering
\includegraphics[width=0.48\textwidth]{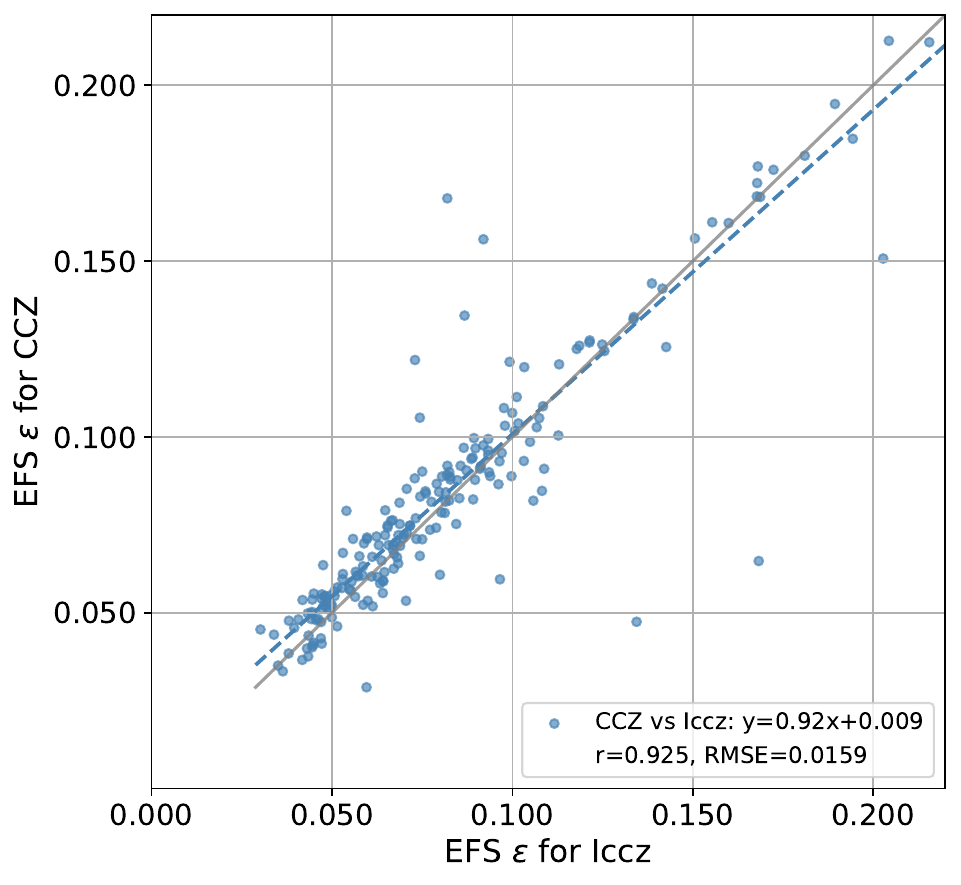}
\caption{\justifying EFS estimates for CCZ versus $\Iccz$
on \texttt{ibm\_kingston} across all valid qubit triples
($\alpha=0.92$, $r=0.925$, $\mathrm{RMSE}=0.0159$).
The close agreement for the large majority of triples
confirms the expected dominance of diagonal noise; the few
outliers ($\approx 3$--$4\%$) reveal measurable non-diagonal
error structure made detectable by the $R_z(\pm\pi/4)$
redistribution.}
\label{fig:efs_ccz_vs_iccz}
\end{figure}

%7.15
Having validated EFS against independent IRB on $\Iccz$
and CZ$(0,2)$, we now apply EFS directly to CCZ on the
same qubit triples. Beyond this IRB validation on Clifford
targets, the comparison between CCZ and $\Iccz$ provides a
controlled stress test of the estimator: the virtual
$R_z(\pm\pi/4)$ rotations leave diagonal error channels
unchanged while redistributing coherent $X/Y$ components
(Sec.~\ref{sec:intro}), so the response of EFS to this
redistribution probes a sensitivity that the Clifford
validation alone does not address.
%7.16
EFS(CCZ) is compared with EFS($\Iccz$) from
Sec.~\ref{sec:iccz}, executed sequentially on the same
qubit triples using the same feature set, to deliver the
intended non-Clifford infidelity estimate and to check
for any systematic differences between the two targets.
%7.17
Fig.~\ref{fig:efs_ccz_vs_iccz} shows EFS(CCZ) vs
EFS($\Iccz$) for all valid triples, with slope $\alpha=0.92$,
Pearson correlation $r=0.925$, and $\mathrm{RMSE}=0.0159$.

%7.18
The two estimates agree closely for the large majority of
triples, confirming the expected null hypothesis that, under
diagonal-dominated noise, $\varepsilon(\mathrm{CCZ}) \approx
\varepsilon(\Iccz)$: the virtual $R_z(\pm\pi/4)$ rotations
commute with $Z$-type, relaxation ($T_1$), and dephasing
($T_2$) errors, leaving the infidelity unchanged. A small
fraction of triples, approximately 3--4\%, depart from this
agreement, revealing measurable non-diagonal ($X/Y$) error
structure that the redistribution makes detectable. This
two-regime outcome --- predominantly diagonal noise with a
minority of triples exhibiting coherent $X/Y$ contributions
--- is consistent with the analysis-mixer role of the EFS
feature pool described in Sec.~\ref{sec:pool}.
%7.19
For reference and completeness, both EFS($\Iccz$) and
EFS(CCZ) are also shown against IRB($\Iccz$) in
Appendix~\ref{sec:appendix_c}
(Fig.~\ref{fig:efs_iccz_ccz_vs_irb_iccz_kingston}).

% ------------------------------------------------------------

%% file: sec_8_A.tex
% sec_8.tex — Conclusion

\section{Conclusion}
\label{sec:conclusion}

%8.1
We have presented an Ensemble Feature Selection (EFS) method
for fast estimation of process infidelity of involutory
multi-qubit gates, including non-Clifford gates for which
standard benchmarking references are unavailable. The method
selects a compact set of circuit measurements from a candidate
pool through offline training on a physically motivated
ensemble of noisy channels, and combines them into a linear
estimator with weights learned by ridge regression.

%8.2
Validated on \texttt{ibm\_kingston} using $\Iccz$ and
CZ$(0,2)$ as Clifford validation benchmarks
with independent IRB, the method achieves an estimation
precision of approximately $0.01$ over a process
infidelity range of $0.02$--$0.20$, using only
$2\Mid=48$ feature circuits.

%8.3
Beyond the IRB validation on Clifford targets, the
comparison between EFS(CCZ) and EFS($\Iccz$) on the same
qubit triples provides a controlled stress test of the
estimator. The virtual $R_z(\pm\pi/4)$ rotations leave
diagonal error channels unchanged while redistributing
coherent $X/Y$ components; the close agreement observed for
the large majority of triples confirms the expected
dominance of diagonal noise, while the small fraction of
outliers, approximately 3--4\%, reveals detectable
non-diagonal contributions. The agreement between EFS and
IRB on $\Iccz$, together with this observed response to the
virtual $R_z(\pm\pi/4)$ redistribution, indicates that the
estimator captures not only diagonal error channels but also
the redistribution of coherent $X/Y$ errors characteristic of
the non-Clifford implementation.

%8.4
The ensemble design is an explicit and tunable component
of the protocol. When an independent IRB reference is available for a Clifford gate, the agreement EFS$\approx$IRB serves a dual purpose --- validating the estimator for non-Clifford targets on the same device, and indicating that the ensemble captures the dominant features of the processor noise. Iterative adjustment of the ensemble
spectral weights to improve this agreement thus constitutes
both a calibration procedure for non-Clifford benchmarking
and a principled approach to hardware noise characterization.

%8.5
While not aimed at ultra-high-precision characterization, the method provides rapid diagnostic feedback using fewer than 50 circuits, enabling efficient identification of poorly performing qubit triples. Such information can directly support calibration, error-mitigation, and quantum error-correction workflows. 

%8.6 
Each component can be improved independently, and further gains are expected from refined noise ensembles, enhanced feature-pool design, and more advanced learning strategies. We therefore view the present results as evidence that resource-efficient, hardware-informed characterization of non-Clifford gates is feasible in practice.

%% file: sec_8_Ack.tex
\acknowledgements
This research is supported by the Bulgarian national plan for recovery and resilience, contract BG-RRP-2.004-0008-C01 (SUMMIT: Sofia University Marking Momentum for Innovation and Technological Transfer), project number 3.1.4. and by the European Union’s Horizon Europe research and innovation program under Grant Agreement No. 101046968 (BRISQ).

%% file: sec_9_A.tex
\clearpage
\appendix

\section{Gate Circuits}
\label{sec:appendix_circ}

\begin{widetext}
\noindent\begin{minipage}[t]{0.48\textwidth}
\vspace{0pt}

%A1.1
This appendix shows the particular transpiled benchmark
circuits used in the experimental validation of the EFS
estimator. These circuits are not unique; many alternative
realizations could be constructed on the same hardware
topology.

%A1.2
Figure~\ref{fig:ccz_trans_draw} shows the transpiled CCZ
implementation used throughout this work. Removing the
seven virtual $R_z(\pm\pi/4)$ rotations yields the
Clifford circuit $\Iccz$, which can be benchmarked
independently using IRB.

%A1.3
Figure~\ref{fig:cz02_circuit} shows the transpiled
implementation of CZ$(0,2)$. Owing to its lower process
infidelity, this circuit extends the study into a
lower-infidelity regime while retaining a genuinely
three-qubit error structure through the active
participation of the routing qubit.

%A1.4
Both $\Iccz$ and CZ$(0,2)$ are employed as Clifford validation benchmarks for assessing the
performance of the EFS estimator on the non-Clifford
CCZ target.

\vspace{1em}
\centering
\scalebox{0.5}{
\begin{quantikz}[column sep=0.3cm, row sep=0.3cm]
\lstick{$q_0$} & \qw & \qw & \qw & \qw & \qw & \qw & \ctrl{1} & \qw & \qw & \qw & \qw & \qw & \qw & \qw \\
\lstick{$q_1$} & \gate{\sqrt{X}} & \ctrl{1} & \gate{\sqrt{X}} & \ctrl{1} & \gate{\sqrt{X}} & \ctrl{1} & \control{} & \gate{\sqrt{X}} & \ctrl{1} & \gate{\sqrt{X}} & \ctrl{1} & \gate{\sqrt{X}} & \ctrl{1} & \qw \\
\lstick{$q_2$} & \gate{\sqrt{X}} & \ctrl{-1} & \gate{\sqrt{X}} & \ctrl{-1} & \gate{\sqrt{X}} & \ctrl{-1} & \gate{\sqrt{X}} & \qw & \ctrl{-1} & \gate{\sqrt{X}} & \ctrl{-1} & \gate{\sqrt{X}} & \ctrl{-1} & \qw
\end{quantikz}
}

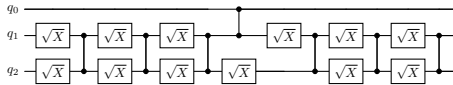
\captionof{figure}{\justifying Transpiled circuit for CZ$(0,2)$ on a
three-qubit topology, containing 12 $\sqrt{X}$ and 7
native CZ gates (depth 13). Qubit $q_1$ acts as a routing
qubit and participates actively in the decomposition,
making CZ$(0,2)$ a genuinely three-qubit operation from
the perspective of its error structure. The circuit serves
as a Clifford validation benchmark, providing
validation of the EFS estimator in a lower
process-infidelity regime.}
\label{fig:cz02_circuit}
\end{minipage}\hfill\begin{minipage}[t]{0.48\textwidth}
\vspace{0pt}
\centering
\scalebox{0.45}{
\begin{tabular}{l}

\begin{quantikz}
\lstick{$q_0$} & \qw & \qw & \qw & \qw & \qw & \qw & \qw & \rstick{\small >>} \\
\lstick{$q_1$} & \qw & \qw & \qw & \qw & \qw & \qw & \qw & \rstick{\small >>} \\
\lstick{$q_2$}
& \gate{R_Z(\pi/2)} & \gate{\sqrt{X}} & \gate{R_Z(\pi/2)} & \gate{R_Z(\pi/2)}
& \gate{\sqrt{X}} & \gate{R_Z(\pi/2)} & \gate{R_Z(\pi/2)} & \rstick{\small >>}
\end{quantikz}

\\[15mm]

\begin{quantikz}
\lstick{$q_0$} & \qw & \qw & \qw & \qw & \qw & \qw & \qw & \rstick{\small >>} \\
\lstick{$q_1$} & \qw & \qw & \ctrl{1} & \gate{\sqrt{X}} & \qw & \qw & \qw & \rstick{\small >>} \\
\lstick{$q_2$}
& \gate{\sqrt{X}} & \gate{R_Z(\pi/2)} & \ctrl{-1} & \gate{R_Z(\pi/2)}
& \gate{\sqrt{X}} & \gate{R_Z(\pi/2)} & \RZfourth{-\pi/4} & \rstick{\small >>}
\end{quantikz}

\\[15mm]

\begin{quantikz}
\lstick{$q_0$}
& \qw & \qw & \qw & \qw & \qw & \qw & \qw & \qw & \qw & \ctrl{1} & \qw & \rstick{\small >>} \\
\lstick{$q_1$}
& \qw & \qw & \qw & \qw & \ctrl{1} & \gate{\sqrt{X}} & \ctrl{1} & \gate{\sqrt{X}}
& \ctrl{1} & \ctrl{-1} & \gate{R_Z(\pi/2)} & \rstick{\small >>} \\
\lstick{$q_2$}
& \gate{R_Z(\pi/2)} & \gate{\sqrt{X}} & \gate{R_Z(\pi/2)} & \gate{\sqrt{X}}
& \ctrl{-1} & \gate{\sqrt{X}} & \ctrl{-1} & \gate{\sqrt{X}} & \ctrl{-1} & \qw & \qw
& \rstick{\small >>}
\end{quantikz}

\\[15mm]

\begin{quantikz}
\lstick{$q_0$} & \qw & \qw & \qw & \qw & \qw & \qw & \qw & \rstick{\small >>} \\
\lstick{$q_1$}
& \gate{\sqrt{X}} & \gate{R_Z(\pi/2)} & \RZfourth{\pi/4} & \gate{R_Z(\pi/2)}
& \gate{\sqrt{X}} & \gate{R_Z(\pi/2)} & \ctrl{1} & \rstick{\small >>} \\
\lstick{$q_2$} & \qw & \qw & \qw & \qw & \qw & \qw & \ctrl{-1} & \rstick{\small >>}
\end{quantikz}

\\[15mm]

\begin{quantikz}
\lstick{$q_0$} & \qw & \qw & \qw & \qw & \qw & \qw & \rstick{\small >>} \\
\lstick{$q_1$}
& \gate{R_Z(\pi/2)} & \gate{\sqrt{X}} & \gate{R_Z(\pi/2)} & \RZfourth{-\pi/4}
& \gate{R_Z(\pi/2)} & \gate{\sqrt{X}} & \rstick{\small >>} \\
\lstick{$q_2$}
& \RZfourth{\pi/4} & \gate{R_Z(\pi/2)} & \gate{\sqrt{X}} & \gate{R_Z(\pi/2)}
& \gate{\sqrt{X}} & \qw & \rstick{\small >>}
\end{quantikz}

\\[15mm]

\begin{quantikz}
\lstick{$q_0$} & \qw & \ctrl{1} & \qw & \qw & \qw & \qw & \qw & \rstick{\small >>} \\
\lstick{$q_1$}
& \gate{R_Z(\pi/2)} & \ctrl{-1} & \gate{R_Z(\pi/2)} & \gate{\sqrt{X}}
& \gate{R_Z(\pi/2)} & \RZfourth{\pi/4} & \gate{R_Z(\pi/2)} & \rstick{\small >>} \\
\lstick{$q_2$} & \qw & \qw & \qw & \qw & \qw & \qw & \qw & \rstick{\small >>}
\end{quantikz}

\\[15mm]

\begin{quantikz}
\lstick{$q_0$}
& \qw & \qw & \qw & \qw & \qw & \qw & \qw & \qw & \qw & \qw & \rstick{\small >>} \\
\lstick{$q_1$}
& \gate{\sqrt{X}} & \gate{R_Z(\pi/2)} & \gate{R_Z(\pi/2)} & \gate{\sqrt{X}}
& \gate{R_Z(\pi/2)} & \gate{\sqrt{X}} & \ctrl{1} & \gate{\sqrt{X}} & \ctrl{1}
& \gate{\sqrt{X}} & \rstick{\small >>} \\
\lstick{$q_2$}
& \qw & \qw & \qw & \qw & \qw & \qw & \ctrl{-1} & \gate{\sqrt{X}} & \ctrl{-1}
& \gate{\sqrt{X}} & \rstick{\small >>}
\end{quantikz}

\\[15mm]

\begin{quantikz}
\lstick{$q_0$}
& \qw & \ctrl{1} & \RZfourth{\pi/4} & \qw & \qw & \qw & \qw & \qw & \rstick{\small >>} \\
\lstick{$q_1$}
& \ctrl{1} & \ctrl{-1} & \gate{R_Z(\pi/2)} & \gate{\sqrt{X}} & \gate{R_Z(\pi/2)}
& \RZfourth{-\pi/4} & \gate{R_Z(\pi/2)} & \gate{\sqrt{X}} & \rstick{\small >>} \\
\lstick{$q_2$}
& \ctrl{-1} & \qw & \qw & \qw & \qw & \qw & \qw & \qw & \rstick{\small >>}
\end{quantikz}

\\[15mm]

\begin{quantikz}
\lstick{$q_0$} & \qw & \ctrl{1} & \qw & \qw & \qw \\
\lstick{$q_1$}
& \gate{R_Z(\pi/2)} & \ctrl{-1} & \gate{R_Z(\pi/2)} & \gate{\sqrt{X}} & \gate{R_Z(\pi/2)} \\
\lstick{$q_2$} & \qw & \qw & \qw & \qw & \qw
\end{quantikz}

\end{tabular}
}
\captionof{figure}{\justifying Transpiled CCZ circuit on a three-qubit
topology, containing 28 $\sqrt{X}$, 12 CZ, 32 virtual
$R_z(\pi/2)$, and 7 virtual $R_z(\pm\pi/4)$ operations.
Removing the seven $R_z(\pm\pi/4)$ rotations (highlighted
in orange) yields $\Iccz$, which serves as a Clifford
validation benchmark chosen to approximate the
coherent and decoherent noise structure of the original
CCZ circuit.}
\label{fig:ccz_trans_draw}
\end{minipage}
\end{widetext}

%% file: sec_9_B.tex
%\newpage
\onecolumngrid

\section{Extended Feature Set ($M=36$)}
\label{sec:appendix_b}
\twocolumngrid

\begin{widetext}
\begin{minipage}[t]{0.48\textwidth}
\vspace{0pt}

The scatter plot and feature table for $\Mid=36$ are provided
here to support the claim in Sec.~\ref{sec:results} that
increasing beyond $\Mid=24$ yields no substantial improvement.

\vspace{1em}
\centering
\includegraphics[width=\textwidth]{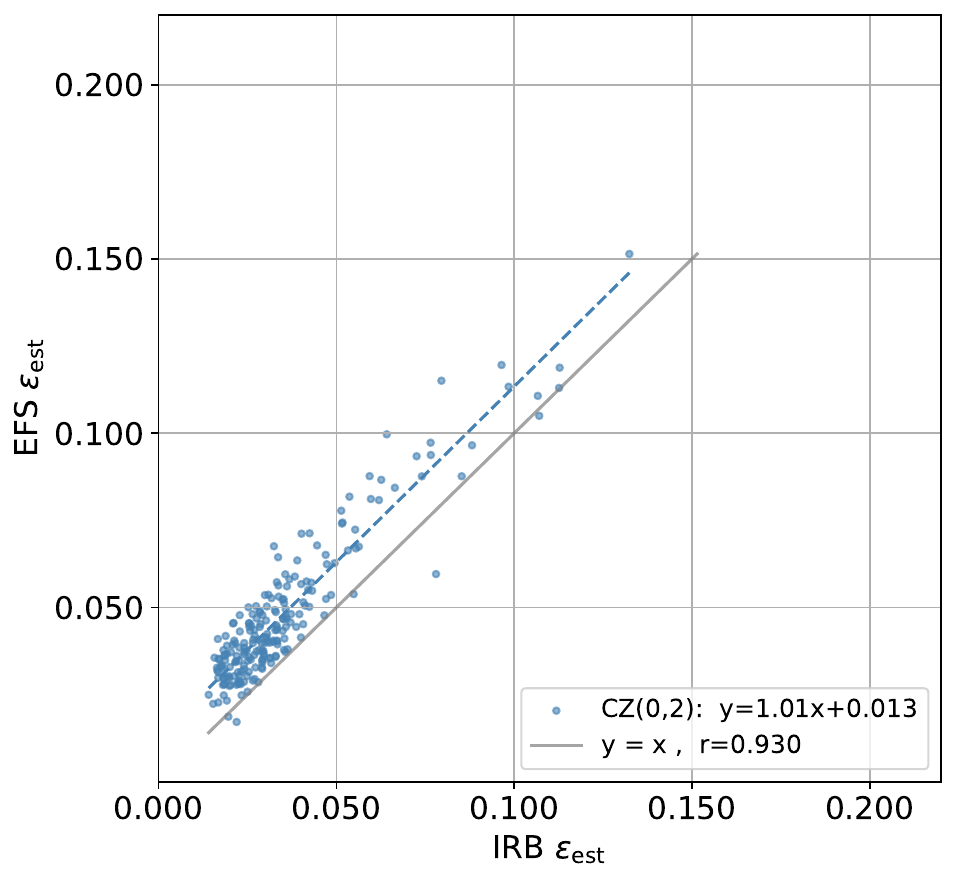}
\captionof{figure}{\justifying EFS vs IRB for CZ$(0,2)$ on
\texttt{ibm\_kingston} with $\Mid=36$. The improvement over
$\Mid=24$ is marginal ($\alpha=1.008$, $r=0.930$,
$\mathrm{RMSE}_{\mathrm{cal}}=0.0078$ vs $0.0081$),
confirming practical saturation at $\Mid\approx 24$.}
\label{fig:cz02_kingston_36}
\end{minipage}
\hfill
\begin{minipage}[t]{0.48\textwidth}
\vspace{0pt}
\centering
\small
\setlength{\tabcolsep}{4pt}
\renewcommand{\arraystretch}{1.05}
\captionof{table}{\justifying Selected EFS features for
CZ$(0,2)$ with $\Mid=36$.}
\label{tab:efs_features_36}
\begin{tabular}{c c c c c}
\hline
$m$ & $\psi_m$ & $\phi_m\{q_2,q_1,q_0\}$ & $L_m$ & $w_m \times 10^2$ \\
\hline
 1 & $|{+}\rangle|{+}\rangle|{+}\rangle$ & $(0,-\pi/2,0)$            & 4 & $-1.78$ \\
 2 & $|1\rangle|0\rangle|0\rangle$        & $(0,0,\pi)$               & 4 & $-1.55$ \\
 3 & $|i\rangle|{+}\rangle|i\rangle$      & $(+\pi/2,0,-\pi/2)$       & 4 & $-3.24$ \\
 4 & $|{+}\rangle|{+}\rangle|i\rangle$    & $(+\pi/2,\pi,-\pi/2)$     & 8 & $+1.83$ \\
 5 & $|0\rangle|0\rangle|1\rangle$        & $(0,0,0)$                 & 4 & $-1.56$ \\
 6 & $|{+}\rangle|{+}\rangle|{+}\rangle$  & $(\pi,\pi,-\pi/2)$        & 4 & $-2.07$ \\
 7 & $|1\rangle|1\rangle|1\rangle$        & $(\pi,0,\pi)$             & 4 & $-1.36$ \\
 8 & $|i\rangle|i\rangle|i\rangle$        & $(-\pi/2,+\pi/2,0)$       & 4 & $-2.20$ \\
 9 & $|1\rangle|1\rangle|1\rangle$        & $(+\pi/2,\pi,+\pi/2)$     & 4 & $-1.87$ \\
10 & $|{+}\rangle|i\rangle|i\rangle$      & $(+\pi/2,0,-\pi/2)$       & 4 & $-2.16$ \\
11 & $|1\rangle|1\rangle|0\rangle$        & $(0,+\pi/2,-\pi/2)$       & 4 & $-1.77$ \\
12 & $|i\rangle|{+}\rangle|{+}\rangle$    & $(-\pi/2,+\pi/2,\pi)$     & 4 & $-2.02$ \\
13 & $|1\rangle|0\rangle|1\rangle$        & $(\pi,-\pi/2,0)$          & 4 & $-1.77$ \\
14 & $|0\rangle|0\rangle|0\rangle$        & $(+\pi/2,-\pi/2,-\pi/2)$  & 4 & $-1.91$ \\
15 & $|0\rangle|0\rangle|1\rangle$        & $(+\pi/2,-\pi/2,\pi)$     & 4 & $-5.32$ \\
16 & $|0\rangle|1\rangle|0\rangle$        & $(-\pi/2,0,0)$            & 4 & $-1.42$ \\
17 & $|{+}\rangle|{+}\rangle|{+}\rangle$  & $(-\pi/2,\pi,+\pi/2)$     & 4 & $-2.82$ \\
18 & $|0\rangle|i\rangle|{+}\rangle$      & $(-\pi/2,\pi,+\pi/2)$     & 8 & $+0.78$ \\
19 & $|0\rangle|i\rangle|{+}\rangle$      & $(0,\pi,0)$               & 8 & $-1.03$ \\
20 & $|{+}\rangle|i\rangle|i\rangle$      & $(+\pi/2,-\pi/2,\pi)$     & 4 & $-1.93$ \\
21 & $|{+}\rangle|0\rangle|{+}\rangle$    & $(\pi,0,\pi)$             & 4 & $+1.64$ \\
22 & $|{+}\rangle|1\rangle|i\rangle$      & $(\pi,-\pi/2,-\pi/2)$     & 4 & $+1.42$ \\
23 & $|1\rangle|1\rangle|0\rangle$        & $(+\pi/2,\pi,+\pi/2)$     & 4 & $-2.29$ \\
24 & $|i\rangle|{+}\rangle|i\rangle$      & $(+\pi/2,0,-\pi/2)$       & 8 & $+1.29$ \\
25 & $|{+}\rangle|i\rangle|i\rangle$      & $(+\pi/2,\pi,+\pi/2)$     & 4 & $-0.93$ \\
26 & $|i\rangle|i\rangle|1\rangle$        & $(-\pi/2,-\pi/2,-\pi/2)$  & 4 & $+1.49$ \\
27 & $|{+}\rangle|i\rangle|0\rangle$      & $(-\pi/2,+\pi/2,+\pi/2)$  & 4 & $+1.41$ \\
28 & $|0\rangle|0\rangle|0\rangle$        & $(+\pi/2,+\pi/2,\pi)$     & 4 & $-3.96$ \\
29 & $|{+}\rangle|{+}\rangle|i\rangle$    & $(0,0,0)$                 & 8 & $-0.81$ \\
30 & $|0\rangle|i\rangle|i\rangle$        & $(\pi,+\pi/2,0)$          & 4 & $+1.96$ \\
31 & $|1\rangle|{+}\rangle|{+}\rangle$    & $(-\pi/2,\pi,+\pi/2)$     & 4 & $+1.85$ \\
32 & $|i\rangle|i\rangle|i\rangle$        & $(\pi,-\pi/2,0)$          & 4 & $-1.68$ \\
33 & $|1\rangle|{+}\rangle|i\rangle$      & $(\pi,\pi,\pi)$           & 8 & $-0.45$ \\
34 & $|{+}\rangle|i\rangle|{+}\rangle$    & $(-\pi/2,-\pi/2,+\pi/2)$  & 8 & $+1.36$ \\
35 & $|{+}\rangle|i\rangle|{+}\rangle$    & $(\pi,+\pi/2,+\pi/2)$     & 4 & $-1.78$ \\
36 & $|i\rangle|{+}\rangle|i\rangle$      & $(+\pi/2,\pi,\pi)$        & 4 & $-1.43$ \\
\hline
\end{tabular}
\end{minipage}
\end{widetext}

%% file: sec_9_C.tex
%\newpage
\clearpage
\onecolumngrid
\section{Additional Figure}
\label{sec:appendix_c}
\twocolumngrid

\begin{figure}[t]
\centering
\includegraphics[width=0.48\textwidth]{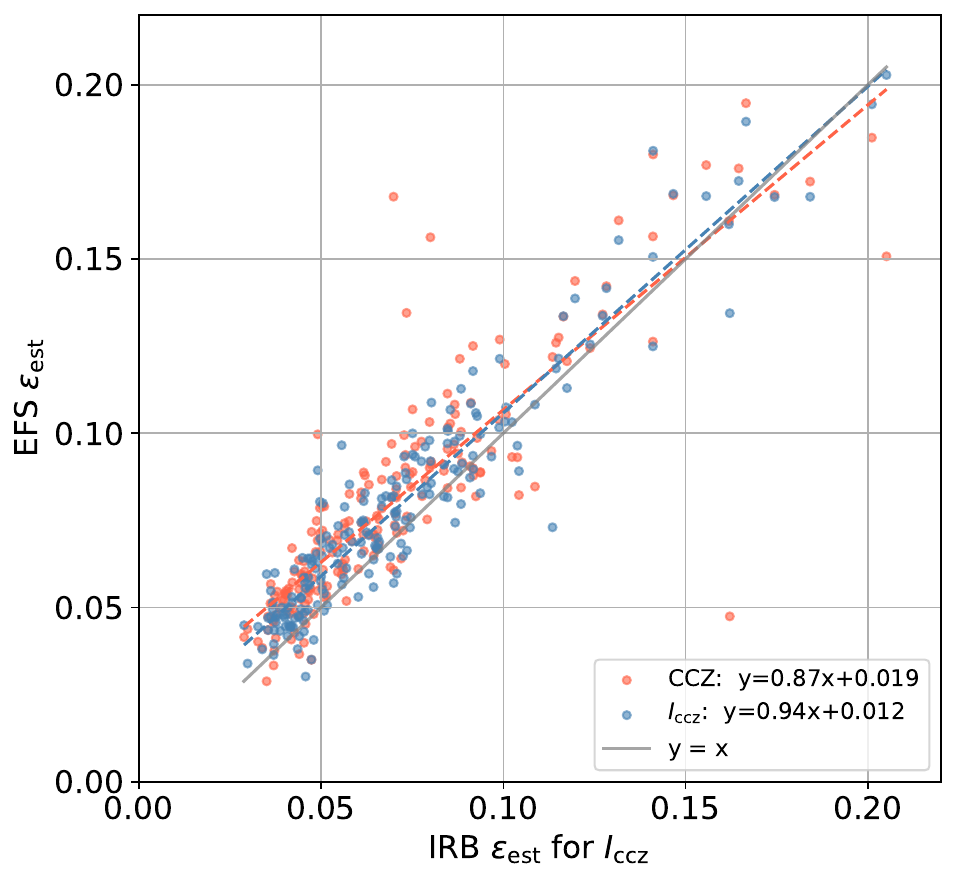}
\caption{\justifying EFS($\Iccz$) and EFS(CCZ) versus
IRB($\Iccz$) on \texttt{ibm\_kingston}, derived from the
same data as Fig.~\ref{fig:efs_ccz_vs_iccz}.
$\Iccz$: slope $0.94$, CCZ: slope $0.87$.}
\label{fig:efs_iccz_ccz_vs_irb_iccz_kingston}
\end{figure}

%C.1
Fig.~\ref{fig:efs_iccz_ccz_vs_irb_iccz_kingston} is derived from
the same data as Fig.~\ref{fig:efs_ccz_vs_iccz} and is
included for reference and completeness, showing how both
EFS($\Iccz$) and EFS(CCZ) compare against IRB($\Iccz$)
on the same qubit triples.

